\documentclass[aps,prl,epsfig,graphics,showpacs,twocolumn,floatfix,mathbbm,a4paper]{revtex4}
\newcommand{\id}{{\sf 1 \hspace{-0.3ex} \rule{0.1ex}{1.52ex}\rule[-.01ex]{0.3ex}{0.1ex}}}

\usepackage{amsmath,amsfonts,amssymb,graphics,graphicx,epsfig,color,times,bbm}
\usepackage{theorem}

\begin{document}
\bibliographystyle{apsrev}
\title{Distributed implementation of standard oracle operators}
\author{Anthony Chefles}
\affiliation{Quantum Information
Processing Group, Hewlett-Packard Laboratories, Filton Road, Stoke
Gifford, Bristol BS34 8QZ UK }
\begin{abstract}
The standard oracle operator corresponding to a function $f$ is a
unitary operator that computes this function coherently, i.e. it
maintains superpositions.  This operator acts on a bipartite system,
where the subsystems are the input and output registers. In
distributed quantum computation, these subsystems may be spatially
separated, in which case we will be interested in its classical and
entangling capacities. For an arbitrary function $f$, we show that
the unidirectional classical and entangling capacities of this
operator are $\log_{2}(n_{f})$ bits/ebits, where $n_{f}$ is the
number of different values this function can take. An optimal
procedure for bidirectional classical communication with a standard
oracle operator corresponding to a permutation on $\mathbb{Z}_{M}$
is given.  The bidirectional classical capacity of such an operator
is found to be $2{\log}_{2}(M)$ bits.   The proofs of these
capacities are facilitated by an optimal distributed protocol for
the implementation of an arbitrary standard oracle operator.
\end{abstract}
\maketitle

The rapidly developing field of quantum information science has
yielded many new concepts in communications and computation, which
have led to major applications such as quantum cryptography and fast
quantum algorithms \cite{NC}.  In quantum, as in classical
information processing, situations involving spatially separated
parties are of particular interest. It is therefore necessary to
develop the theory of distributed quantum information processing
\cite{GC,CGB,EJPP,CLP,KC,BHLS,LHL,VNM}. Here, we consider quantum
systems whose component subsystems are possessed by a number of
spatially-separated parties.  These subsystems cannot interact
directly, so the effect of an interaction must be brought about
using only local quantum operations and non-local resources. The
non-local resources needed to implement an arbitrary quantum
operation in this manner are classical communication channels and
shared entangled states.

One particularly important quantum operation within the context of
quantum information processing is the standard oracle operator. This
operator is a key building block for quantum algorithms. Generally
speaking, an oracle operator is a unitary operator that computes a
function.  The key difference between oracle operators and classical
methods of computation is that the former, being linear quantum
mechanical operators, maintain superpositions.  A superposition of
different values of the independent variable, which we denote by
$x$, will then evolve into a superposition of the corresponding
values of $f(x)$, giving rise to the well-known and important
phenomenon of quantum parallelism. The standard oracle operator is a
convenient oracle operator which can be used to compute an arbitrary
function \cite{KKVB,CKTST}.  However, it has different registers for
$x$ and $f(x)$ and in a distributed setting it is natural to
consider these to be spatially separated.

In this Letter, we investigate numerous aspects of the distributed
implementation of standard oracle operators.  We consider both the
minimum entanglement and classical communication resources, in both
directions, required for this implementation and also the
corresponding capacities, which relate to the fact that it is
possible to use such an operator to send classical information and
create entangled states. It is important to determine the values of
these quantities for the following reasons. Regarding the minimal
resources necessary for the distributed implementation of the
standard oracle operator, it is highly desirable to use classical
communication and, even more so, shared entanglement, as efficiently
as possible when implementing distributed quantum operations.
Concerning the capacities, it is important to have knowledge of
these quantities in circumstances where we are able to perform this
operation and wish to use it to create entangled states or transmit
classical information.

The main result of this Letter is that for an arbitrary function
$f$, all six minimum implementation resources and capacities are
equal to ${\log}_{2}(n_{f})$ bits/ebits, where $n_{f}$ is the number
of different values this function can take.  In the course of this
investigation, we provide optimal protocols for entanglement
creation and classical communication using an arbitrary standard
oracle operator, indeed also for bidirectional classical
communication when the function $f$ is a permutation.  We also give
an optimal protocol for the distributed implementation of an
arbitrary standard oracle operator.

Let us set the scene by reviewing the main properties of standard
oracle operators. Let $M,N$ be arbitrary finite integers ${\geq}1$.
Consider ${\cal F}_{MN}$, the set of functions from
$\mathbb{Z}_{M}{\mapsto}\mathbb{Z}_{N}$. Let $A$ and $B$ be quantum
systems with $M$- and $N$-dimensional Hilbert spaces ${\cal H}_{M}$
and ${\cal H}_{N}$.  These systems are taken to be spatially
separated and in the possession of corresponding parties Alice and
Bob.  To each $f{\in}{\cal F}_{MN}$ there corresponds a unitary
standard oracle operator on ${\cal H}_{M}{\otimes}{\cal H}_{N}$:
\begin{equation}
\label{oracleaction}
U_{f}|x{\rangle}_{A}{\otimes}|y{\rangle}_{B}=|x{\rangle}_{A}{\otimes}|y{\oplus}f(x){\rangle}_{B}.
\end{equation}
$A$ and $B$ may be referred to as the control and target systems
respectively.  In Eq. \eqref{oracleaction}, ${\oplus}$ denotes
addition modulo $N$. Also, $x{\in}\mathbb{Z}_{M},
y{\in}\mathbb{Z}_{N}$ and $\{|x{\rangle}\}$ is an orthonormal basis
set for ${\cal H}_{M}$, likewise with $\{|y{\rangle}\}$ and ${\cal
H}_{N}$. These  are the computational basis sets for both systems.
There are $N^{M}$ functions in ${\cal F}_{MN}$, so there are $N^{M}$
associated standard oracle operators $U_{f}$.

To proceed, let us partition $\mathbb{Z}_{M}$ into subsets
corresponding to different values of $f(x)$.  Let $n_{f}$ be the
number of different values that $f(x)$ can take.  Clearly,
$n_{f}{\leq}M,N$.  Let $f_{j}$, where
$j{\in}\{0,{\ldots},n_{f}-1\}$, be the possible values of $f(x)$.
We also define $S_{j}{\subset}\mathbb{Z}_{M}$ to be the set of
values of $x$ for which $f(x)=f_{j}$ and denote by
$P_{j}=\sum_{x{\in}S_{j}}|x{\rangle}{\langle}x|$ the projector onto
the subspace spanned by the states $|x{\rangle}$ for $x{\in}S_{j}$.
Finally, let $K_{j}$ be the cardinality of $S_{j}$. Clearly, $K_{j}$
is the rank of $P_{j}$.  It is a simple matter to prove that $U_{f}$
can be written in the form \cite{CKTST}
\begin{equation}
\label{USchmidt}
U_{f}=\sum_{j=0}^{n_{f}-1}P_{jA}{\otimes}(e^{-if_{j}{\Phi}_{N}})_{
B},
\end{equation}
where we use the $N$-dimensional Pegg-Barnett phase operator
\begin{equation}
{\Phi}_{N}=\sum_{n{\in}\mathbb{Z}_{N}}\frac{2{\pi}n}{N}|{\phi}_{Nn}{\rangle}{\langle}{\phi}_{Nn}|,
\end{equation}
whose eigenstates are the $N$-dimensional Pegg-Barnett phase states
$|{\phi}_{Nn}{\rangle}=N^{-1/2}\sum_{y{\in}\mathbb{Z}_{N}}e^{\frac{2{\pi}iny}{N}}|y{\rangle}$
\cite{PB}. These states form an orthonormal basis for ${\cal H}_{N}$
which is conjugate to the computational basis $\{|y{\rangle}\}$.
One can readily verify that
\begin{equation}
\label{plusone}
e^{-i{\Phi}_{N}}|y{\rangle}=|y{\oplus}1{\rangle}\;\;{\forall\;\;}y{\in}\mathbb{Z}_{N}.
\end{equation}
We note that Eq. \eqref{USchmidt} gives an operator Schmidt
decomposition of $U_{f}$, where the related Schmidt operator sets
are $\{P_{j}/\sqrt{K_{j}}\}$ and
$\{e^{-if_{j}{\Phi}_{N}}/\sqrt{N}\}$. These are orthonormal sets
with respect to the Hilbert-Schmidt inner product
${\langle}A,B{\rangle}=\mathrm{Tr}(A^{\dagger}B)$. The Schmidt
coefficients are $\sqrt{NK_{j}}$ and the Schmidt rank of $U_{f}$,
denoted by $\mathrm{Sch}(U_{f})$, is equal to $n_{f}$.

In a distributed setting, any type of non-local resource that can be
created by a quantum operation must also be consumed in order to
perform the operation.   For a bipartite quantum operation, there
are three such resources: shared entanglement $E$ and classical
communication in the Alice$\rightarrow$Bob and Bob$\rightarrow$Alice
directions, which we shall denote by $C_{\rightarrow}$ and
$C_{\leftarrow}$ respectively.  We shall use the subscripts $R$ and
$C$ to denote, respectively, the minimum of the corresponding
resource required to perform a quantum operation and the capacity of
the operation corresponding to this resource. The entangling
capacity is the maximum amount of entanglement that the operation
can create.  The classical capacity, in a given direction, is the
maximum amount of classical information that the operation can be
used to send in that direction.

A fundamental result in quantum information theory is that, for any
bipartite unitary operator $U$, each capacity cannot exceed the
amount of the corresponding resource that must be consumed
\cite{EJPP}. We therefore have the following inequalities:
\begin{eqnarray}
\label{in1}
E_{R}(U)&{\geq}&E_{C}(U), \\
\label{in2}
C_{R{\rightarrow}}(U)&{\geq}&C_{C{\rightarrow}}(U), \\
\label{in3} C_{R{\leftarrow}}(U)&{\geq}&C_{C{\leftarrow}}(U).
\end{eqnarray}
There is a further capacity to consider, the bidirectional classical
capacity $C_{C{\leftrightarrow}}(U)$. This is the maximum total
amount of classical information that Alice and Bob can send to each
other with one use of the quantum operation. Since the
unidirectional classical capacities are optimised for transmission
in their associated directions, we have
\begin{equation}
\label{bcc}
C_{C{\leftrightarrow}}(U){\leq}C_{C{\rightarrow}}(U)+C_{C{\leftarrow}}(U).
\end{equation}
We shall now obtain, for an arbitrary standard oracle operator
$U_{f}$, lower bounds on the entangling and unidirectional classical
capacities $E_{C}(U_{f})$, $C_{C{\rightarrow}}(U_{f})$ and
$C_{C{\leftarrow}}(U_{f})$. We begin by examining entanglement
creation. Consider some arbitrary but fixed $x_{j}{\in}S_{j}$, for
each $j{\in}\{0,{\ldots},n_{f}-1\}$. Suppose that $A$ and $B$ are
initially prepared in the product state
\begin{equation}
|{\chi}{\rangle}=\left(\frac{1}{\sqrt{n_{f}}}\sum_{j=0}^{n_{f}-1}|x_{j}{\rangle}_{A}\right){\otimes}|0{\rangle}_{B},
\end{equation}
where $|0{\rangle}$ is the zeroth computational basis state in
${\cal H}_{N}$.  Acting upon this state with $U_{f}$ gives
\begin{equation}
U_{f}|{\chi}{\rangle}=\frac{1}{\sqrt{n_{f}}}\sum_{j=0}^{n_{f}-1}|x_{j}{\rangle}_{A}{\otimes}|f_{j}{\rangle}_{B}.
\end{equation}
This is a maximally entangled state with Schmidt rank $n_{f}$,
having ${\log}_{2}(n_{f})$ ebits of entanglement.  We conclude that
\begin{equation}
\label{ebound} E_{C}(U_{f}){\geq}{\log}_{2}(n_{f}).
\end{equation}
Let us now show that Alice and Bob can send each other
${\log}_{2}(n_{f})$ classical bits using $U_{f}$. That Alice can
send Bob ${\log}_{2}(n_{f})$ bits is almost trivially demonstrated.
Let $r{\in}\{0,{\ldots},n_{f}-1\}$ be the classical message she
wishes to send to Bob. She prepares $A$ in the state
$|x_{r}{\rangle}$. Meanwhile, Bob prepares $B$ in the state
$|0{\rangle}$. The oracle operator $U_{f}$ then acts on these
systems, giving rise to the state
$|x_{r}{\rangle}_{A}{\otimes}|f_{r}{\rangle}_{B}$. Bob can
subsequently perform a computational basis measurement to reveal
$f_{r}$ and hence $r$, Alice's ${\log}_{2}(n_{f})$ bit message.

For Bob to send the same amount of classical information to Alice,
the two parties can use the following entangled state:
\begin{equation}
\label{psi}
|{\psi}{\rangle}=\frac{1}{\sqrt{n_{f}}}\sum_{j=0}^{n_{f}-1}|x_{j}{\rangle}_{A}{\otimes}|N{\ominus}f_{j}{\rangle}_{B},
\end{equation}
where ${\ominus}$ denotes subtraction modulo $N$. Bob wishes to send
the value of $s{\in}\{0,{\ldots},n_{f}-1\}$ to Alice. To encode his
chosen value of $s$ in the above state, he makes use of a unitary
phase shift operator $G$ acting on ${\cal H}_{N}$ which is defined
through
\begin{equation}
G|N{\ominus}f_{j}{\rangle}=e^{\frac{2{\pi}ij}{n_{f}}}|N{\ominus}f_{j}{\rangle}.
\end{equation}
His encoding of $s$ is performed through the transformation
$|{\psi}{\rangle}{\mapsto}|{\psi}_{s}{\rangle}=({\id}_{A}{\otimes}G^{s}_{B})|{\psi}{\rangle}$,
giving
\begin{equation}
|{\psi}_{s}{\rangle}=\frac{1}{\sqrt{n_{f}}}\sum_{j=0}^{n_{f}-1}e^{\frac{2{\pi}ijs}{n_{f}}}|x_{j}{\rangle}_{A}{\otimes}|N{\ominus}f_{j}{\rangle}_{B}.
\end{equation}
The oracle operator $U_{f}$ is then applied, resulting in the state
\begin{equation}
U_{f}|{\psi}_{s}{\rangle}=\left(\frac{1}{\sqrt{n_{f}}}\sum_{j=0}^{n_{f}-1}e^{\frac{2{\pi}ijs}{n_{f}}}|x_{j}{\rangle}_{A}\right){\otimes}|N{\rangle}_{B}.
\end{equation}
 The states
inside the parentheses, indexed by $s$, are orthonormal and can be
perfectly discriminated by Alice.  Doing so enables her to read
Bob's ${\log}_{2}(n_{f})$ bit message $s$.  The existence of these
classical communication protocols implies that
\begin{equation}
\label{ccbound}
C_{C{\rightarrow}}(U_{f}),C_{C{\leftarrow}}(U_{f}){\geq}{\log}_{2}(n_{f}).
\end{equation}
Let us now consider simultaneous, bidirectional classical
communication.  Here we will see that, when $f$ is permutation from
$\mathbb{Z}_{M}{\mapsto}\mathbb{Z}_{M}$, the above protocol can be
modified to enable Alice and Bob to send to each other
${\log}_{2}(n_{f})={\log}_{2}(M)$ classical bits simultaneously. Let
$f:\mathbb{Z}_{M}{\mapsto}\mathbb{Z}_{M}$ be a permutation of degree
$M$.  We begin with the state
\begin{equation}
|{\Psi}{\rangle}=\frac{1}{\sqrt{M}}\sum_{x{\in}\mathbb{Z}_{M}}|x{\rangle}_{A}{\otimes}|M{\ominus}x{\rangle}_{B},
\end{equation}
which resembles the state $|{\psi}{\rangle}$ in Eq. \eqref{psi}.
Here, ${\oplus}/{\ominus}$ denotes addition/subtraction modulo $M$.
Alice encodes her message $r{\in}\mathbb{Z}_{M}$ with the unitary
transformation $|x{\rangle}{\mapsto}|f^{-1}(x{\oplus}r){\rangle}$ on
$A$. Again, Bob encodes his message $s$ with a unitary phase shift
on $B$, here
$|M{\ominus}x{\rangle}{\mapsto}e^{\frac{2{\pi}isx}{M}}|M{\ominus}x{\rangle}$
where $s{\in}\mathbb{Z}_{M}$.  The total state transformation is
$|{\Psi}{\rangle}{\mapsto}|{\Psi}_{rs}{\rangle}$, where
\begin{equation}
|{\Psi}_{rs}{\rangle}=\frac{1}{\sqrt{M}}\sum_{x{\in}\mathbb{Z}_{M}}e^{\frac{2{\pi}isx}{M}}|f^{-1}(x{\oplus}r){\rangle}_{A}{\otimes}|M{\ominus}x{\rangle}_{B}.
\end{equation}
The corresponding standard oracle operator $U_{f}$ is then applied,
which results in the transformation
\begin{equation}
U_{f}|{\Psi}_{rs}{\rangle}=\frac{1}{\sqrt{M}}\sum_{x{\in}\mathbb{Z}_{M}}e^{\frac{2{\pi}isx}{M}}|f^{-1}(x{\oplus}r){\rangle}_{A}{\otimes}|M{\oplus}r{\rangle}_{B}.
\end{equation}
Alice and Bob are now able to read each other's messages.  For the
sake of clarity, let Alice now invert her earlier unitary
transformation on $A$ and Bob perform the unitary transformation
$\sum_{r'{\in}\mathbb{Z}_{M}}|r'{\rangle}{\langle}M{\oplus}r'|$ on
$B$. This results in the state
\begin{equation}
\left(\frac{1}{\sqrt{M}}\sum_{x{\in}\mathbb{Z}_{M}}e^{\frac{2{\pi}isx}{M}}|x{\rangle}_{A}\right){\otimes}|r{\rangle}_{B}.
\end{equation}
The states of $A$ are the orthonormal eigenstates of ${\Phi}_{M}$
indexed by $s$.  These states are perfectly distinguishable by
Alice,  as are the states $|r{\rangle}$ by Bob. Discrimination among
these states enables Alice and Bob to read each other's
${\log}_{2}(M)$ bit messages.  We therefore conclude, for a standard
oracle operator corresponding to a permutation of degree $M$, that
the bidirectional classical capacity satisfies
\begin{equation}
\label{bccbound}
C_{C{\leftrightarrow}}(U_{f}){\geq}2{\log}_{2}(M).
\end{equation}

Having obtained lower bounds on the entangling and classical
capacities for a standard oracle operator, we now obtain upper
bounds on the corresponding minimum resources for its distributed
implementation.  We will now show that
\begin{equation}
\label{rbound}
E_{R}(U_{f}),C_{R{\rightarrow}}(U_{f}),C_{R{\leftarrow}}(U_{f}){\leq}{\log}_{2}(n_{f}),
\end{equation}
by describing an explicit protocol that uses ${\log}_{2}(n_{f})$
ebits of entanglement and the same number of classical bits in each
direction to perform the distributed implementation of a standard
oracle operator.  We begin with an arbitrary initial state of
systems $A$ and $B$, which may be written in the form
\begin{equation}
|{\Phi}{\rangle}=\sum_{m{\in}\mathbb{Z}_{M} \atop
n{\in}\mathbb{Z}_{N}}c_{mn}|m{\rangle}_{A}{\otimes}|n{\rangle}_{B}.
\end{equation}
In addition to $A$ and $B$, Alice and Bob have respective ancillas
$a$ and $b$.  Their Hilbert spaces can be described in the following
way. Let us define ${\cal H}_{f}$ as the $n_{f}$-dimensional
subspace of ${\cal H}_{M}$ spanned by the states $|x_{j}{\rangle}$
for $j{\in}\{0,{\ldots},n_{f}-1\}$. Then the Hilbert spaces of $a$
and $b$ are copies of ${\cal H}_{f}$. The two ancillas are initially
prepared in the maximally entangled state
$n_{f}^{\-1/2}\sum_{j=0}^{n_{f}-1}|x_{j}{\rangle}_{a}{\otimes}|x_{j}{\rangle}_{b}$,
which has ${\log}_{2}(n_{f})$ ebits of entanglement.  The total
initial state is therefore
\begin{equation}
|{\Phi}_{0}{\rangle}=\frac{1}{\sqrt{n_{f}}}\sum_{j=0}^{n_{f}-1}\sum_{m{\in}\mathbb{Z}_{M}
\atop
n{\in}\mathbb{Z}_{N}}c_{mn}|m{\rangle}_{A}{\otimes}|x_{j}{\rangle}_{a}{\otimes}|n{\rangle}_{B}{\otimes}|x_{j}{\rangle}_{b}.
\end{equation}
Our protocol can be described in the following way:\\

{\noindent \em Step 1:} Alice applies the following unitary operator
to $Aa$:
\begin{equation}
{\Omega}=\sum_{k=0}^{n_{f}-1}P_{k}{\otimes}V_{k}.
\end{equation}
Here, $V_{k}$ is a unitary operator on ${\cal H}_{f}$ which acts as
$V_{k}|x_{j}{\rangle}=|x_{j\bar{\oplus}k}{\rangle}$, where
throughout, $\bar{\oplus}/\bar{\ominus}$ denotes
addition/subtraction modulo $n_{f}$. The state transformation
effected by this operator is
$|{\Phi}_{0}{\rangle}{\mapsto}|{\Phi}_{1}{\rangle}$, where
\begin{equation}
|{\Phi}_{1}{\rangle}=\frac{1}{\sqrt{n_{f}}}\sum_{j,k=0}^{n_{f}-1}\sum_{m{\in}S_{k}
\atop
n{\in}\mathbb{Z}_{N}}c_{mn}|m{\rangle}_{A}{\otimes}|x_{j\bar{\oplus}k}{\rangle}_{a}{\otimes}|n{\rangle}_{B}{\otimes}|x_{j}{\rangle}_{b}.
\end{equation}
{\noindent \em Step 2:} Alice performs a computational basis
measurement on $a$, getting result $x_{r}$ for some
$r{\in}\{0,{\ldots},n_{f}-1\}$.  This results in the state
transformation $|{\Phi}_{1}{\rangle}{\mapsto}|{\Phi}_{2r}{\rangle}$
where
\begin{equation}
|{\Phi}_{2r}{\rangle}=\sum_{k=0}^{n_{f}-1}\sum_{m{\in}S_{k} \atop
n{\in}\mathbb{Z}_{N}}c_{mn}|m{\rangle}_{A}{\otimes}|x_{r}{\rangle}_{a}{\otimes}|n{\rangle}_{B}{\otimes}|x_{r\bar{\ominus}k}{\rangle}_{b}.
\end{equation}

{\noindent \em Step 3:} Alice communicates the value of $r$ to Bob,
thus sending him ${\log}_{2}(n_{f})$ classical bits. With his
knowledge of $r$, Bob performs the unitary transformation
$|x_{r\bar{\ominus}k}{\rangle}{\mapsto}|x_{k}{\rangle}$ on $b$,
resulting in the total state transformation
$|{\Phi}_{2r}{\rangle}{\mapsto}|{\Phi}_{3r}{\rangle}$ where
\begin{equation}
|{\Phi}_{3r}{\rangle}=\sum_{k=0}^{n_{f}-1}\sum_{m{\in}S_{k} \atop
n{\in}\mathbb{Z}_{N}}c_{mn}|m{\rangle}_{A}{\otimes}|x_{r}{\rangle}_{a}{\otimes}|n{\rangle}_{B}{\otimes}|x_{k}{\rangle}_{b}.
\end{equation}
{\noindent \em Step 4:} Bob now performs the unitary transformation
\begin{equation}
|n{\rangle}_{B}{\otimes}|x_{k}{\rangle}_{b}{\mapsto}|n{\oplus}f_{k}{\rangle}_{B}{\otimes}|x_{k}{\rangle}_{b},
\end{equation}
where ${\oplus}$ denotes addition modulo $N$.  This transformation
is effectively the oracle operator $U_{f}$, with $b$ and $B$ being
the control and target systems respectively and the state of the
control system is restricted to the subspace ${\cal H}_{f}$ of
${\cal H}_{M}$.  The gives
$|{\Phi}_{3r}{\rangle}{\mapsto}|{\Phi}_{4r}{\rangle}$ where
\begin{equation}
|{\Phi}_{4r}{\rangle}=\frac{1}{\sqrt{n_{f}}}\sum_{k=0}^{n_{f}-1}\sum_{m{\in}S_{k}
\atop
n{\in}\mathbb{Z}_{N}}c_{mn}|m{\rangle}_{A}{\otimes}|x_{r}{\rangle}_{a}{\otimes}|n{\oplus}f_{k}{\rangle}_{B}{\otimes}|x_{k}{\rangle}_{b}.
\end{equation}
{\noindent \em Step 5:} Bob performs a discrete Fourier transform
 on the $b$ system whose effect is
 $|x_{k}{\rangle}{\mapsto}\frac{1}{\sqrt{n_{f}}}\sum_{s=0}^{n_{f}-1}e^{\frac{2{\pi}iks}{n_{f}}}|x_{s}{\rangle}$,
resulting in the total state transformation
$|{\Phi}_{4r}{\rangle}{\mapsto}|{\Phi}_{5r}{\rangle}$, where
\begin{eqnarray}
|{\Phi}_{5r}{\rangle}&=&\frac{1}{\sqrt{n_{f}}}\sum_{k,s=0}^{n_{f}-1}\sum_{m{\in}S_{k}
\atop n{\in}\mathbb{Z}_{N}}c_{mn}e^{\frac{2{\pi}iks}{n_{f}}}
|m{\rangle}_{A}{\otimes}|x_{r}{\rangle}_{a} \nonumber
\\ &{\otimes}&|n{\oplus}f_{k}{\rangle}_{B}{\otimes}|x_{s}{\rangle}_{b}.
\end{eqnarray}
{\noindent \em Step 6:} Bob now performs a computational basis
measurement on $b$.  On obtaining the result $x_{s}$, where
$s{\in}\{0,{\ldots},n_{f}-1\}$, the total state is transformed as
\begin{eqnarray}
|{\Phi}_{5r}{\rangle}{\mapsto}|{\Phi}_{6rs}{\rangle}&=&\sum_{k=0}^{n_{f}-1}\sum_{m{\in}S_{k}
\atop n{\in}\mathbb{Z}_{N}}c_{mn}e^{\frac{2{\pi}iks}{n_{f}}} |m
{\rangle}_{A}{\otimes}|x_{r}{\rangle}_{a} \nonumber
\\ &{\otimes}&|n{\oplus}f_{k}{\rangle}_{B}{\otimes}|x_{s}{\rangle}_{b}
\end{eqnarray}
and he communicates the value of $s$ to Alice.  This requires him to
send her ${\log}_{2}(n_{f})$ bits of classical information.\\

{\noindent \em Step 7:} Alice now uses the degenerate but unitary
phase shift operator
$T=\sum_{k=0}^{n_{f}-1}e^{\frac{-2{\pi}ik}{n_{f}}}P_{k}$. Knowing
$s$, she applies the operator $T^{s}$ to $A$.
 This results in the transformation
 $|{\Phi}_{6rs}{\rangle}{\mapsto}|{\Phi}_{7rs}{\rangle}$, where
\begin{eqnarray}
|{\Phi}_{7rs}{\rangle}&=&\sum_{k=0}^{n_{f}-1}\sum_{m{\in}S_{k} \atop
n{\in}\mathbb{Z}_{N}}c_{mn}
|m{\rangle}_{A}{\otimes}|x_{r}{\rangle}_{a}{\otimes}|n{\oplus}f_{k}{\rangle}_{B}{\otimes}|x_{s}{\rangle}_{b}\nonumber
\\
&=&(U_{f}|{\Phi}{\rangle})_{AB}{\otimes}|x_{r}{\rangle}_{a}{\otimes}|x_{s}{\rangle}_{b},
\end{eqnarray}
which is the desired transformation of the state of $AB$.  The
existence of this protocol for the distributed implementation of the
standard oracle operator $U_{f}$, with the specified resources
together with the lower capacity bounds in \eqref{ebound},
\eqref{ccbound} and inequalities \eqref{in1}, \eqref{in2} and
\eqref{in3}, establishes that all six quantities in these latter
inequalities are equal to ${\log}_{2}(n_{f})$ \cite{footnote}.  We
also see from \eqref{bcc} that when $f$ is a permutation of degree
$M$, the bidirectional classical capacity
$C_{C{\leftrightarrow}}(U_{f})$ is equal to $2{\log}_{2}(n_{f})$
bits and that the bidirectional classical communication protocol we
described is optimal.

There are several points to be made about this distributed protocol.
Firstly, it generalises earlier work on the distributed
implementation of the CNOT gate \cite{GC,EJPP,CLP}. In fact, this
unitary gate is the standard oracle operator corresponding to the
one-bit identity function.   Our protocol has interesting security
properties. The actual classical data that Alice and Bob send to
each other consists of random measurement results.  It follows that
if they wish to use $U_{f}$ to send classical information to each
other, this will be concealed from an eavesdropper listening to
their classical transmissions. Also, we see that in step 4, Bob
effectively implements the oracle locally. Only this step makes
reference to the details of the function $f$, which even Alice
doesn't have to know for the successful implementation of $U_{f}$.
The details of $f$ will also be concealed from a potential
eavesdropper on the classical transmissions.

We also point out that this protocol simplifies when $f$ is a
permutation on $\mathbb{Z}_{M}$.  When this is so, $M=N=n_{f}$ and
all four quantum systems have identical Hilbert spaces.  The
projectors $P_{k}$ have rank-one and project onto all of the
computational basis states in ${\cal H}_{M}$.  One further curious
property of permutations is the ease with which their standard
oracle operators can be seen to be locally equivalent.  Kashefi et
al. \cite{KKVB} noted that for any permutation $f$ on
$\mathbb{Z}_{M}$, one can define the unitary minimal oracle operator
$Q_{f}=\sum_{x{\in}\mathbb{Z}_{M}}|f(x){\rangle}{\langle}x|$, which
is related to $U_{f}$ through $
U_{f}=(Q^{\dagger}_{f}{\otimes}{\id}_{M})U_{ID}(Q_{f}{\otimes}{\id}_{M})$.
Here, $U_{ID}$ is the standard oracle operator corresponding to the
$\mathbb{Z}_{M}{\mapsto}\mathbb{Z}_{M}$ identity function.  All
standard oracle operators for permutations of degree $M$ are
therefore interconvertible with local unitary operations.  It
follows that the minimum non-local resources to implement these
operators and their corresponding capacities are equal.

To conclude, we have studied numerous aspects of the distributed
implementation of standard oracle operators.  These  arise
frequently in the context of quantum algorithms and the results
presented here will be useful in relation to distributed quantum
computation.  It is also to be expected that the methods used to
establish the minimum non-local implementation resources and
capacities of standard oracle operators will be useful in a more
general context. In particular, the optimal distributed protocol for
standard oracle operators has the potential to be modified for more
general unitary operators.

{\noindent \em Acknowledgements}: I would like to thank Timothy
Spiller and Bill Munro for helpful discussions and comments on this
manuscript and also Martin Plenio for many enjoyable conversations
about this topic. Part of this work was carried out at NICT, Tokyo
and I would like to thank Masahide Sasaki and his group for their
hospitality. This work was supported by the EU project QAP.


\begin{thebibliography}{99}
\bibitem{footnote}
It is noteworthy that the entangling capacity $E_{C}(U_{f})$ can
also be deduced from the general inequality
$E_{C}(U){\leq}{\log}_{2}(\mathrm{Sch}(U))$ obtaining by Bennett et
al. \cite{BHLS} for an arbitrary bipartite unitary operator $U$,
together with our entanglement creation protocol and our observation
that $\mathrm{Sch}(U_{f})=n_{f}$.
\end{thebibliography}
\end{document}